\documentclass{iaus}
\usepackage{graphicx}
\usepackage{natbib}

\title[Detection of a stellar halo in NGC 4244] 
{Detection of a stellar halo in NGC 4244}

\author[Seth et al.]   
{A. Seth$^1$, R. de Jong$^2$, J. Dalcanton$^3$, \and the GHOSTS team}

\affiliation{$^1$Harvard-Smithsonian Center for Astrophysics, 60 Garden St., Cambridge, MA 02139 \break email: aseth@cfa.harvard.edu\\[\affilskip]
$^2$Space Telescope Science Institute\\[\affilskip]
$^3$University of Washington}

\pubyear{2007}
\volume{241}  
\pagerange{???}
\date{Jan. 22, 2007}
\jname{Stellar Populations as Building Blocks of Galaxies}
\editors{???, eds.}
\begin{document}

\maketitle

\begin{abstract}
We present the preliminary detection of a stellar halo in the nearby,
edge-on galaxy, NGC 4244.  Using data from the Hubble Space
Telescope's Advanced Camera for Surveys (HST/ACS), we have examined
number counts of red giant branch (RGB) stars along the minor axis of
the galaxy out to 10 kpc, with a limiting surface brightness of $\mu_R
\sim$ 31 mag/arcsec$^2$.  At distances more than 2.5 kpc above the
disk, a second stellar component is clearly visible.  This component
is well fit by either an exponential or a power law.  The best-fit
exponential scale height is similar to the scale length of the galaxy
disk, suggesting that we are indeed detecting a halo component.
Furthermore, the color of the RGB stars above 2.5 kpc are
significantly bluer than those in the disk, suggesting a more
metal-poor population.  Outside the local group, this finding represents one of the
first detections of a resolved stellar halo in a spiral galaxy.
\end{abstract}


The stellar halos of disk galaxies are thought to be made from
merger remnants, and thus trace the earliest epochs of galaxy
formation \citep[e.g.][]{bullock05}.  However, due to their low
surface brightnesses, stellar halos have been detected in only a small
number galaxies, mostly in the Local Group using resolved stars
\citep[e.g.][]{kalirai06,mcconnachie06}. 
Complementing these studies, \citet{zibetti04a} examine the integrated light profiles of $>$1000 stacked edge-on galaxies and show that stellar halos may be a common feature of disk galaxies.


Using HST, we can detect and study stellar halos around nearby
(D$<$10~Mpc) galaxies using resolved RGB stars.  
NGC~4244 is a low mass ($V_{rot} \sim 100$~km/sec, $M_B = -17.7$),
bulgeless spiral, and is one of the nearest edge-on galaxies \citep[D=4.36~Mpc,][]{seth05a}.  The galaxy has smaller dimensions than the
Milky Way with a scale length of 1.8~kpc and a scale height of
$\sim$300~pc.  We have obtained Cycle 14 HST/ACS snapshot images of
seven fields along the principal axes of the disk (Fig.~1,
left panel).  In this contribution, we analyze the distribution and color of
RGB stars along the minor axis.

\noindent {\bf Results: } Figure 2 (left panel) shows the RGB surface
density as a function of height above the
midplane. 
The dip at low disk heights is caused by incompleteness.  At larger disk
heights, a clear exponential drop off is seen that matches the scale
height found by \citet{fry99} using deep ground-based data with a
limiting surface brightness $\mu_R = 27.5$.  We have used the
\citet{fry99} data to calibrate our RGB surface density to $\mu_R$.
Starting $\sim$2.5~kpc above the plane at $\mu_R > 28$, the RGB
surface density starts declining more slowly.  We have fit this 2nd
component to both an exponential and a
power-law, and find that both fit the data well.  The best fitting
exponential has a scale height of $\sim$1.6~kpc, nearly as large as
the scale length of the NGC 4244 disk, while the best-fit power law
falls off as $\sim z^{-2.3}$, similar to the Milky Way halo if seen in
projection.  Either fit suggests that the component we are detecting
is a stellar halo.  This component is {\it much fainter} than the
Milky Way halo; integrating the best-fit exponential assuming a scale
length similar to the disk suggests a mass of only $3\times10^6$~M$_\odot$.

Assuming the 2nd component is old, the color of
the RGB component is dependent on metallicity (Fig.~1, right
panel).  The mean color of RGB stars as a function of disk height 
is shown in the right panel of Fig.~2.  To convert mean color to
metallicity we used synthetic CMDs.  The mean metallicity in the disk
component is [Z]$\sim$-1.1, consistent with the findings of
\citet{seth05b}.  However, above 2.5 kpc, where the 2nd component
dominates, the metallicity drops to $\sim$-1.5.  While this result is
intriguing, it should be viewed with caution, as a better knowledge of
the contaminants' colors is necessary.

\noindent {\bf Conclusions: } We find strong evidence for a second stellar
component along the minor axis of NGC~4244.  Both the morphology and
metallicity of this second component suggest that it is
indeed a stellar halo, similar, but much fainter than that of the
Milky Way.

\begin{figure}
\begin{center}
\includegraphics[height=2.5in]{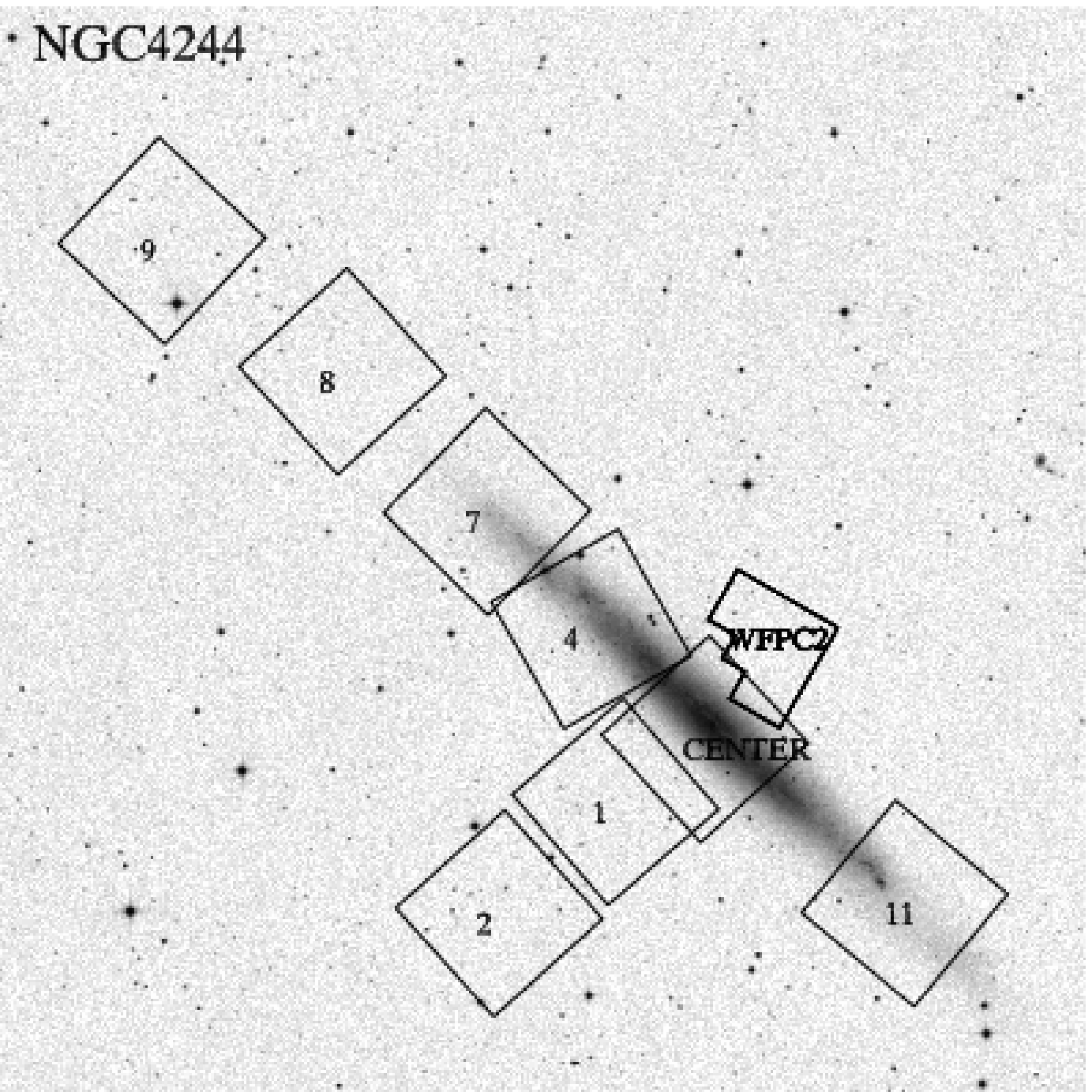}
\includegraphics[height=2.5in]{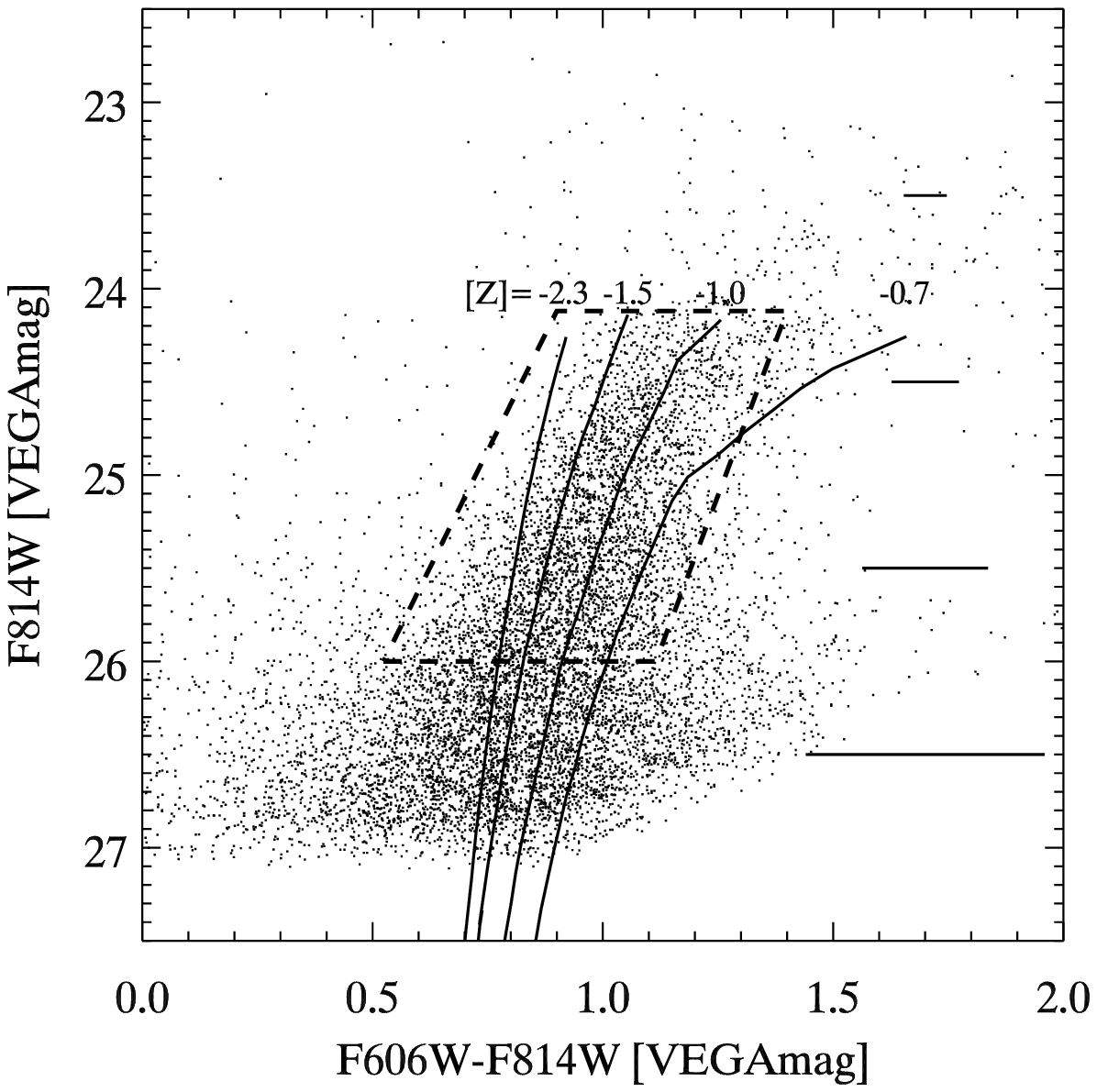}
\end{center}
\caption{{\it Left --} Digitized Sky Survey image of NGC 4244 overlaid
with existing HST imaging. Numbered fields show Cycle 14 GHOSTS data,
here we analyze the three minor axis ACS fields. {\it Right -- }
Color-magnitude diagram of stars more than 1000 pc below the disk.
Overplotted are the box used to study the color of the RGB and 10~Gyr
Padova isochrones at four metallicities.}
\end{figure}

\begin{figure}
\begin{center}
\includegraphics[width=2.8in]{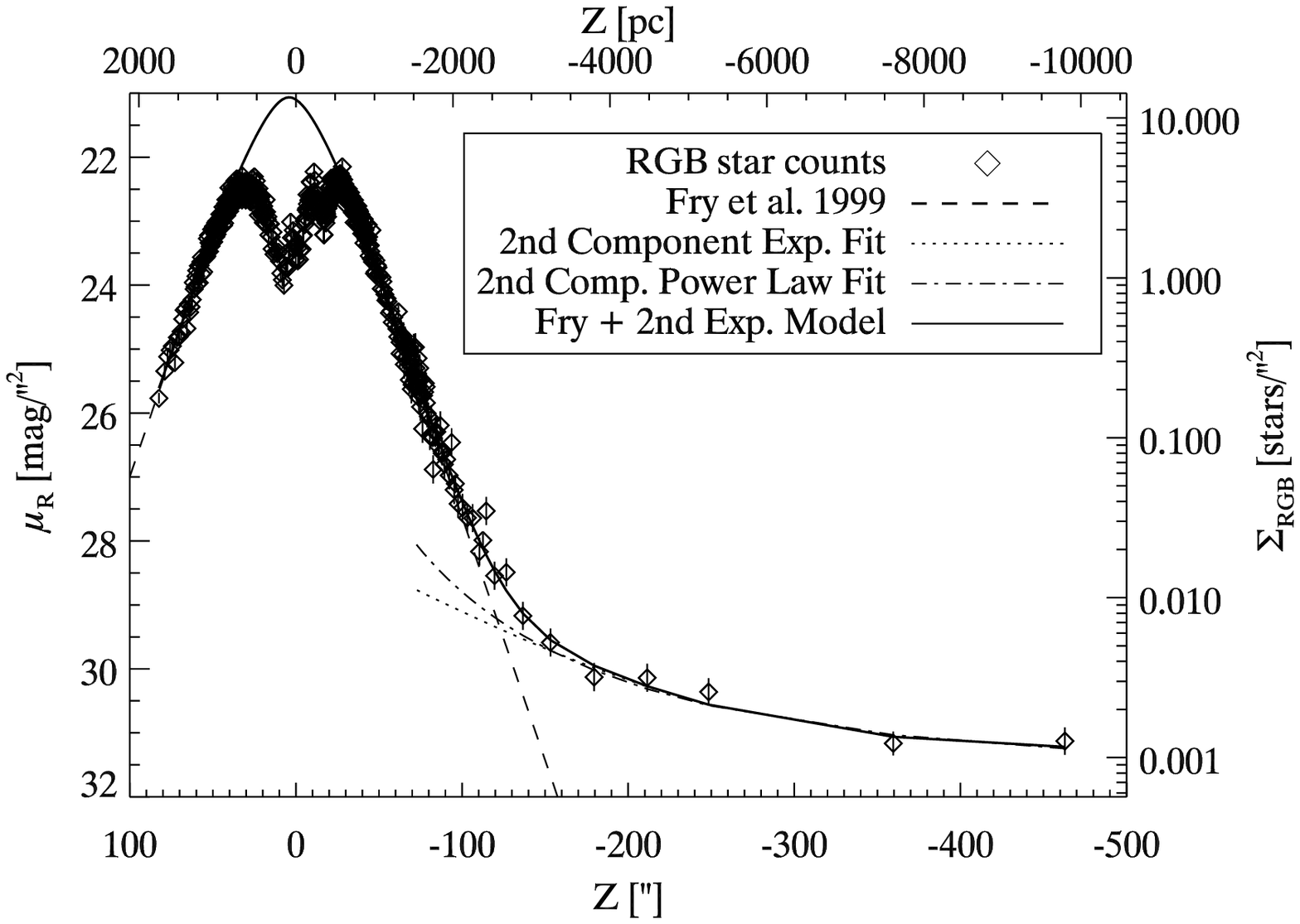}
\includegraphics[width=2.3in]{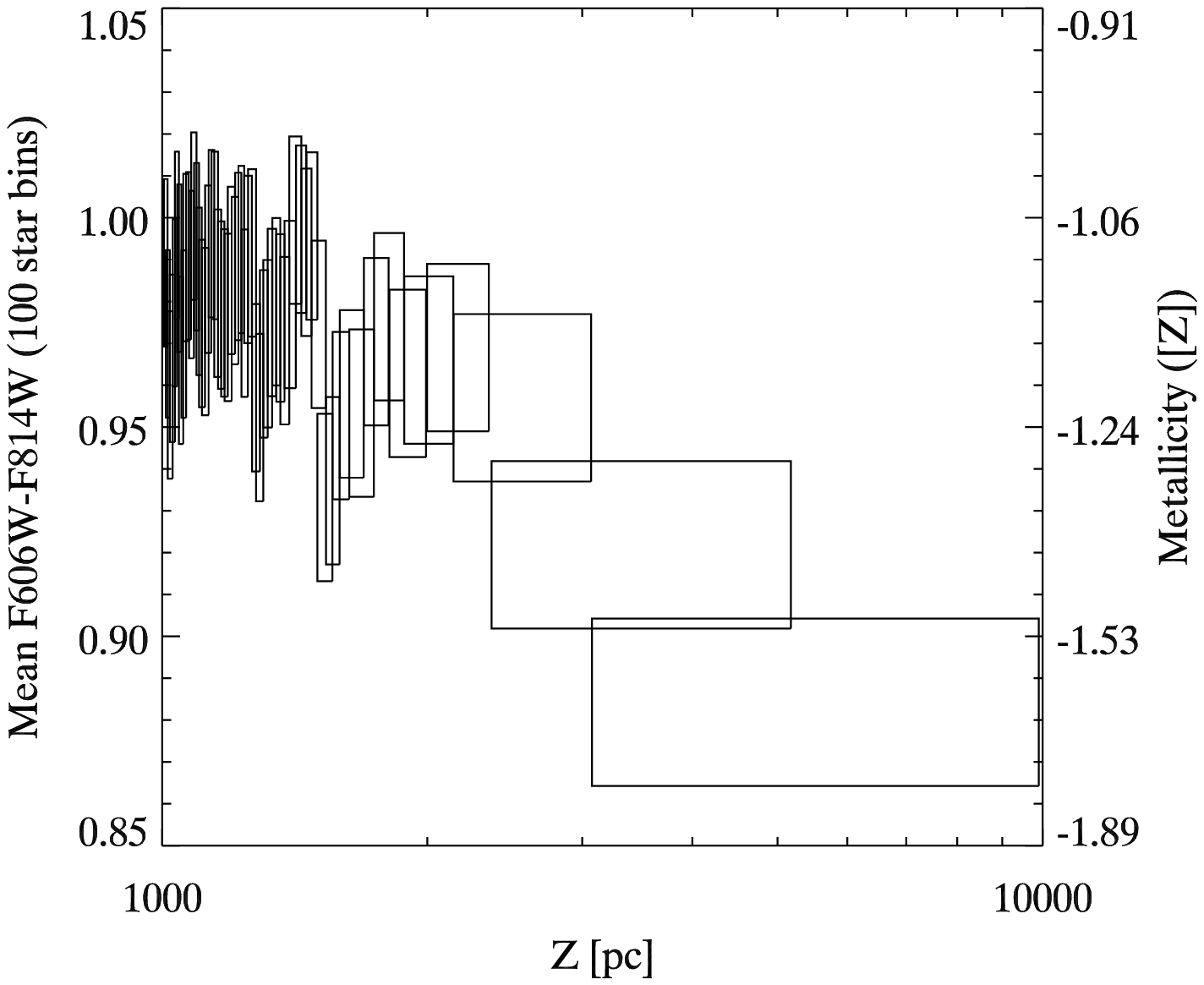}
\end{center}

\caption{{\it Left -- } Minor axis profile of NGC 4244.  Height above
the plane is given in arcseconds (bottom) and parsecs (top).  Diamonds
show our derived RGB surface density.  The dashed line shows the
surface photometry from Fry et al. (1999), while other lines show model
fits.  {\it Right~--~}~Mean color vs. disk height for RGB stars.  Each
box gives the mean color of 100 stars.  Box width indicates the disk
height of stars, and box height gives the error on the mean color.
Equivalent metallicity (right axis) was determined using synthetic
CMDs.}
\end{figure}

\end{document}